\documentclass[%
 aip,
 amsmath,amssymb,
 reprint,%
]{revtex4-1}
\usepackage{graphicx}
\usepackage{amsmath,amssymb}
\usepackage{color}
\usepackage{textcomp}
\usepackage{float}

\begin{document}

\begin{titlepage}
\title{Estimation of the equilibrium free energy for glasses using the Jarzynski equality}
\author{H. A. Vinutha}
\altaffiliation{Corresponding author}
\email{vh163@georgetown.edu}
\affiliation{Department of Physics, Institute for Soft Matter Synthesis and Metrology, Georgetown University, Washington, DC, USA}
\affiliation{Institute of Physics, Chinese Academy of Sciences, Beijing, China}
\affiliation{Yusuf Hamied Department of Chemistry, University of Cambridge, Cambridge, UK}
\author{Daan Frenkel}
\email{df246@cam.ac.uk}
\affiliation{Yusuf Hamied Department of Chemistry, University of Cambridge, Cambridge, UK}

\begin{abstract}
{The free energy of glasses cannot be estimated using thermodynamic integration, as glasses are intrinsically not in equilibrium. 
We present numerical simulations showing that, in contrast, plausible free-energy estimates of a Kob-Andersen glass can be obtained using the  Jarzynski relation. 
 Using the Jarzynski relation, we also compute the chemical potential difference of the two components of this system, and find that, in the glassy regime,  the Jarzynski estimate matches well with the extrapolated value of the supercooled liquid. 
 
  Our findings are of broader interest as they show that the Jarzynski method can be used under conditions where the thermodynamic integration approach, which is normally more accurate, breaks down completely.
Systems where such an approach might be useful are gels and jammed, glassy structures formed by compression. 
}
\end{abstract}
\maketitle
\end{titlepage}
The Helmholtz free energy is one of the most widely used thermodynamic state functions because, for a system of $N$ particles in a fixed volume $V$ at a temperature $T$, the Helmholtz free energy $F$ must be at a minimum when the system has reached  equilibrium.

Computing free energies is therefore important: it allows us to predict the relative stabilities of different states (e.g. phases) of a system.
In thermodynamics, the free-energy difference between two states of a system is related to the reversible work required to bring the system from one state (say $A$) to the other ($B$). 
The work expended during an {\em irreversible} transformation from $A$ to $B$ is larger than the reversible work, and is therefore not a good measure for the free-energy change.

It was therefore a great surprise when Jarzynski~\cite{jarzynski1997equilibrium,jarzynski1997nonequilibrium} showed that  the free-energy difference between two systems can be related to the non-equilibrium work ($W$) required to transform one system into the other in an arbitrarily short time
\begin{equation}\label{eq:JR}
 \exp(-\beta \Delta F) = \overline{\exp(-\beta W(t_s))}\;,
\end{equation}
where   $\beta = \frac{1}{k_BT}$, and the bar over $\exp(-\beta W(t_s))$ denotes averaging over a ``sufficiently large'' number of independent simulations: the term ``sufficiently large'' is necessarily vague because we do not know {\em a priori} how much averaging will be needed for eqn.~\ref{eq:JR} to hold. 

Jarzynski's result stimulated much theoretical work, in particular by Crooks~\cite{crooks1999entropy,crooks2000path,crooks1998nonequilibrium}, who generalized Jarzynski's approach. Moreover, many experiments and simulations have been reported that validated Eqn.~\ref{eq:JR}~ \cite{collin2005verification,douarche2005experimental,toyabe2010experimental}. 

However, in spite of its great conceptual value, it seems that Jarzynski method is not a more accurate or more efficient method to compute free-energy differences than the standard, reversible thermodynamic integration methods~\cite{dellago2014computing,lechner2007efficiency,oberhofer2009efficient,geiger2010optimum}, in situations where such methods can be used.

Here we focus on a problem where thermodynamic integration cannot be used, namely estimating the free energy of a glass.
Glasses are non-equilibrium systems that do not relax on experimentally accessible  time-scales (see e.g. \cite{angell1995formation,debenedetti2001supercooled}). 
It is for this reason that the free energy of a glass cannot be determined by thermodynamic integration: one might even argue that the equilibrium free energy of a glass is an oxymoron. 
However, as has been demonstrated for instance in simulations of polydisperse glasses, \textcolor{black}{it is sometimes possible to equilibrate glassy structures numerically, using so-called ``swap'' moves~\cite{ber171}.  Such an approach will only work for systems where the acceptance of such moves is sufficiently high~\cite{ADSparmar}.
Here we consider the free energy of a glass for which swap moves are inefficient and the approach of ref~\onlinecite{ber171} will not work.}

In this Communication, we show that for glasses where the approach of ref.~\cite{ber171} will not work, the Jarzynski method yields much lower estimates for the free energy of a glass than  the thermodynamic integration method, and thereby provides an interesting method for estimating the free energy of systems that cannot relax to equilibrium. 

Knowledge of the equilibrium free energy of a glass can be of practical use, for instance  for estimating a lower bound to the solubility of a glass. 

There is, however,  a problem: validating our approach against exact free energies is not possible for the widely used Kob-Andersen glass model system that we study~\cite{kob1995testing}.
Hence, as a proxy, we will test which approach yields the lower free-energy estimate, and we will also compare our results with a naive extrapolation of the free energy of a supercooled liquid. 


For equilibrium systems, the free-energy change of a system upon cooling from a temperature ($T_H$) to a lower temperature $T_L$ can be obtained by thermodynamic integration (TI):
\begin{equation}\label{eq:delf_TI}
\beta_L F(T_L) - \beta_H F(T_H) = \int_{\beta_H}^{\beta_L} d\beta \langle U(\beta) \rangle
\end{equation}

We prepare glassy structures by quenching equilibrium liquid configurations from a temperature $T_H$ with cooling rate $C_r$. 
We obtain the free energy of glasses by computing the average work done during the process. 
The relation between cooling and work is discussed  in the Supplementary Material (SM). 
We can then rewrite the Eqn.~\ref{eq:delf_TI}   as follows:
\begin{equation}\label{eq:delf_JR}
\beta_L F(T_L)= \beta_H F(T_H) - \ln \left\langle e^{-[\beta_Lf^{n}(T_L) -\beta_Hf^{n}(T_H)]}\right\rangle_{C_r}
\end{equation}
The difference
\begin{equation}\label{Delta_f}
\beta_Lf^{n}(T_L) -\beta_Hf^{n}(T_H)\equiv \Delta \left(\frac{f^{n}}{T}\right)
\end{equation}
denotes the non-equilibrium work required to change the state of the system within a finite switching time or, equivalently, cooling rate, in a single cooling run.  
In Eq.~\ref{Delta_f}, $n$ stands for the $n^{th}$ cooling run.
 $\Delta \left(\frac{f^{n}}{T}\right)$ is evaluated by computing  the potential energy $E^{n}$ as a function of the inverse temperature $\beta$:
\begin{equation}\label{eq:fb_JR}
 \beta_Lf^{n}(T_L) = \beta_Hf^{n}(T_H) + \int_{\beta_H}^{\beta_L}d\beta\ E^{n}(\beta)
\end{equation}
As explained in the SM, we can recast Eq. \ref{eq:delf_JR} as the effect of scaling the potential energy $U$ rather than the effect of changing the temperature. That is, $\beta F(T)$ for the original temperature but with potential energy function $\lambda U$ has the same value as $\beta^\prime F(T^\prime)$ for the original potential energy function, but temperature $T^\prime = T/\lambda$.
In our calculations, we compute the variation of $\beta F$, as we change the potential energy
at constant $\beta$ for $U$ to $\lambda U$. 
To estimate the free energy difference for the Jarzynsky Relation (JR), we compute
\begin{equation}\label{eq:FU_JR}
 \beta [F(T;\lambda U) - F(T;U)]  = 
 -\ln\langle e^{-\beta\int_1^\lambda d\lambda^\prime \; \overline{U}_{\lambda^\prime} }\rangle ,
 \end{equation}
\textcolor{black}{where $\langle...\rangle$ denotes averaging over all independent slow-cooling runs, whereas $\overline{U}_{\lambda^\prime}$ denotes the average of $U_{\lambda^\prime}$  during a single cooling run.}\\
For our free-energy calculations, we use a well-studied glassy system that can be prepared by slow cooling~\cite{westergren2007silico}, namely the  Kob-Andersen (KA) binary Lennard-Jones model glass former~\cite{kob1995testing,sastry2001relationship,sengupta2011dependence}. 

We simulated $N$=256 ($N_A$=204, $N_B$=52) bi-disperse spheres, 80-20 (A-B) mixture, interacting  via 
$V(r) = 4\epsilon_{\alpha\beta} \left[ \left(\frac{\sigma_{\alpha\beta}}{r}\right)^{12} - \left(\frac{\sigma_{\alpha\beta}}{r}\right)^6 \right] + 4\epsilon_{\alpha\beta} \left[c_0 + c_2 \left(\frac{r}{\sigma_{\alpha\beta}}\right)^2\right]$, for $r_{\alpha\beta} < r_c$, and zero otherwise. 
$r$ denotes the distance between the two particles within in the cutoff distance \cite{sengupta2011dependence}.
We used the standard KA parameter values $\sigma_{AA} = 1.0$, $\sigma_{AB} = 0.8$,$\sigma_{BB} = 0.88$,$r_c = 2.5*\sigma_{\alpha\beta}$, $\epsilon_{AA} = 1.0$, $\epsilon_{AB} = 1.5$, $\epsilon_{BB} = 0.5$. 
$r$ denotes the distance between the two particles within in the cutoff distance \cite{sengupta2011dependence}. $c_0 = 0.01626656, c_2 = -0.001949974$ are fixed by the condition that  the potential and force go  to zero continuously at the cutoff-distance $r_c$. 

 In what follows, all the thermodynamic quantities are expressed in reduced units: 
 $\sigma_{AA}$ is our unit of length, the unit of energy is $\epsilon_{AA}$,  $m_{A}=m_{B}=1$ is defined as the unit of mass, and the reduced temperature is expressed in units $\frac{\epsilon_{AA}}{k_B}$, where $k_B$ is Boltzmann's constant. 
 Below, we report the excess free energy of the system, as the ideal gas part can be computed analytically.

We performed NVT MC simulations at different cooling rates.
Starting with the equilibrium liquid configurations at $T=0.5$ for $N=256$, we performed a stepwise cooling runs to a final temperature of $T$=0.1, in steps of $\Delta T$ = 0.1. 
The ``duration'' of a single cooling step  $\Delta t$ is the number of Monte-Carlo cycles that the system spends at any given temperature.  
We define the cooling rate $C_{\text{r}}$ as  $\Delta T/\Delta t$.  
For instance, for $C_{\text{r}} = 10^{-6}$  we perform $\Delta t = 10^5$ MC cycles at a given temperature. 
Each MC cycle comprises $N$ trial displacement moves. 
To obtain good statistics, we performed $1000$ independent simulation runs for $N=256$ and used a cooling rate $C_r$ = $10^{-6}$.
\\              
\begin{figure}[h!]
\includegraphics[scale=0.4]{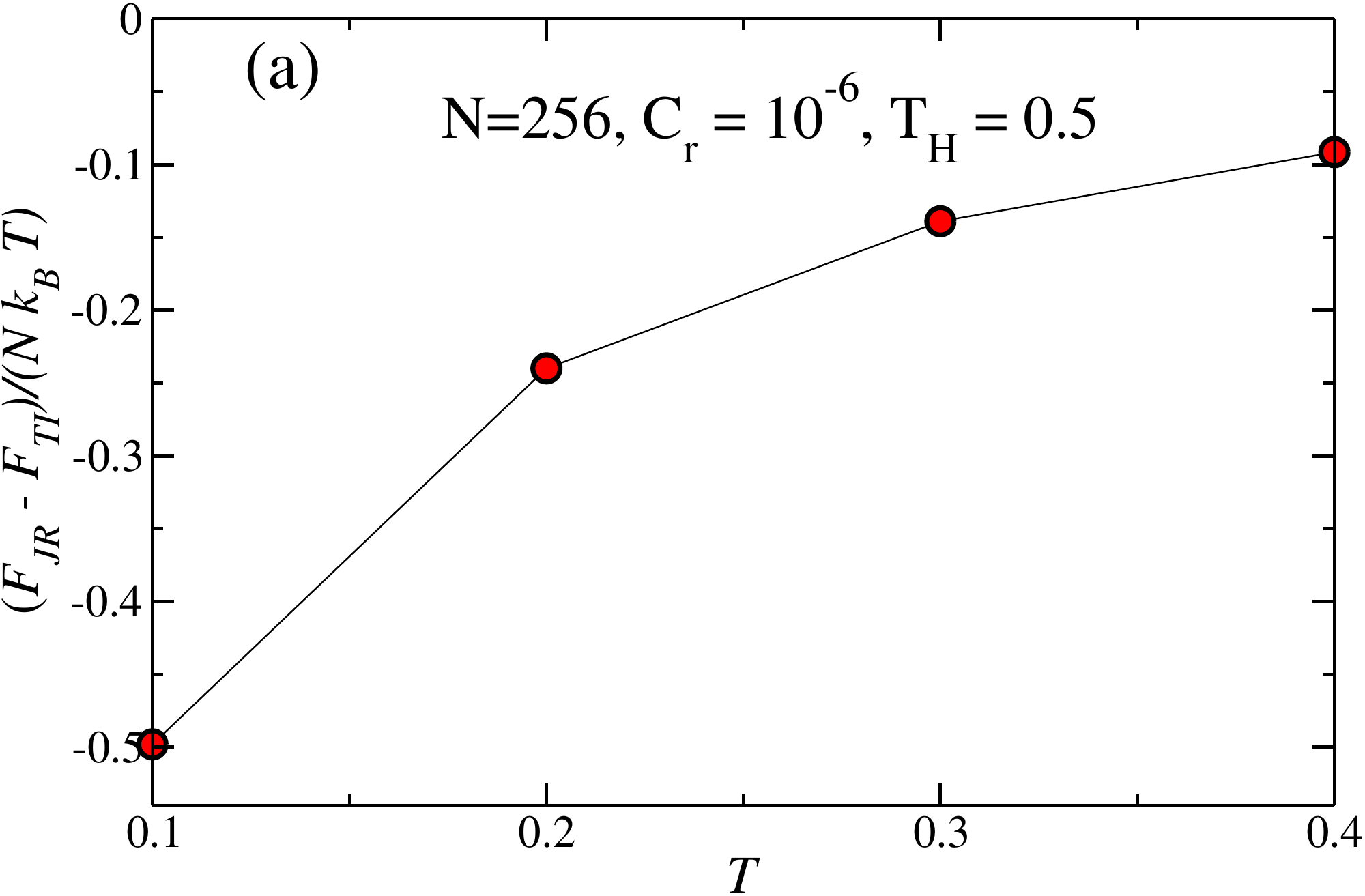}
\includegraphics[scale=0.4,angle=0]{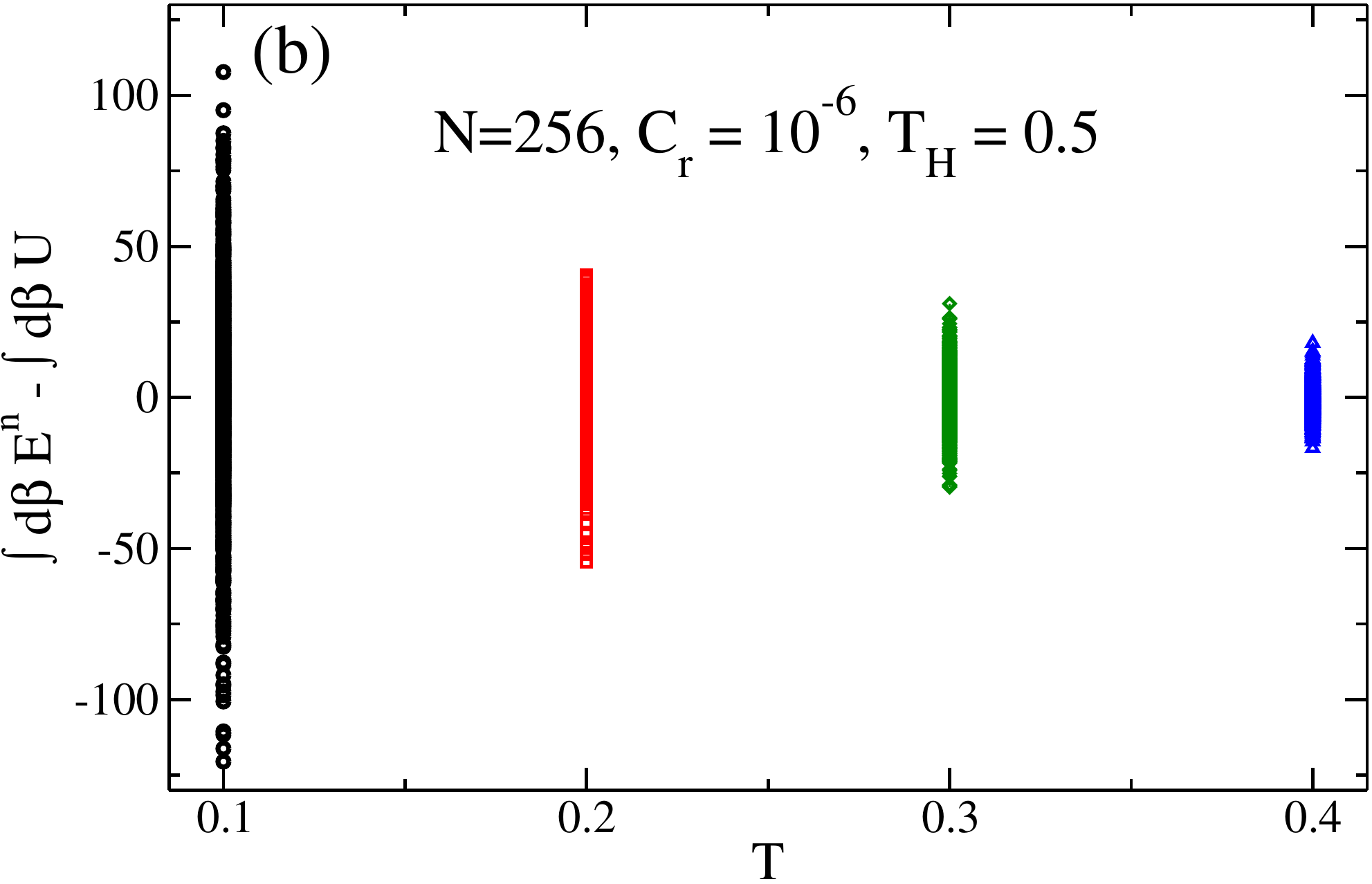}
\caption{\label{Fex_JRTI} {\bf(a)} The difference in the excess free energy obtained using Jarzynski relation (JR) and the thermodynamic integration (TI) method. We show the data for $N$=256,  $C_r$ = $10^{-6}$, the number of samples is $10^3$. \textcolor{black}{Panel {\bf(b)} shows the scatter in the values of the work performed during different cooling runs. 
This wide scatter is typical for glassy system, and would not be observed in systems that equilibrate on the timescale of the simulations. 
It is this scatter that makes it necessary to use the Jarzynski relation, rather than thermodynamic integration (see text).}}
\end{figure}
For each cooling run, we use a quadratic fit form ($U = a_0+a_1 T + a_2 T^2$) to fit the $T$- dependence of The values of $a_0$, $a_1$ and $a_2$ are different for all cooling runs. 
We then use the TI expression,  Eq. \ref{eq:fb_JR}, to compute the difference $\Delta \left(\frac{f^{n}}{T}\right)$ for each run and,   using Eq. \ref{eq:delf_JR} and averaging over different runs, we obtain the JR estimates of $F_{\text{JR}}$, the free energy  of the low-temperature glass, down to a temperature $T$=0.1. 
For the sake of comparison, we also use the TI  method  (Eq.~\ref{eq:delf_TI})  to estimate the free energies of the glasses ($F_{\text{TI}}$)  at temperatures down to $T$=0.1.
Note that we use exactly the same simulation data for the JR and TI estimates (see the SM for more details). 
In Fig. \ref{Fex_JRTI}, we show the difference in the free energy per particle in units of $k_B T$  obtained using TI and JR, for glasses, starting at  $T_H=0.5$.
We note that, as the system is cooled slowly, a cooling run that is started at a higher temperature will traverse all lower temperatures and will therefore, if anything, equilibrate better than a cooling run started a lower temperature. 
We observe that at $T=0.1$, the TI estimate of the free energy per particle is about 0.5 $k_BT$ higher than the estimate obtained using the Jarzynski relation (see Fig. \ref{Fex_JRTI}a). 
\textcolor{black}{Both  free-energy estimates are based on the same raw data: the only difference is how we analyse the data.\\
At low temperatures, the thermodynamic properties of  the system are dominated by low-energy inherent structures. 
If we perform many TI runs some runs will sample lower-energy glassy states than others.
The result of the TI procedure is unweighted average over all these runs, but  in the Jarzynski approach, the results are strongly dominated by those runs that, upon cooling,  end up in low energy inherent structures.
Fig.\ref{Fex_JRTI}b  shows the difference between the work done for different cooling runs and the average work done in the TI runs. 
Of course, as the sampling of low energy states is not exhaustive, in particular at lower temperatures than considered in our study, it is to be expected that even the Jarzynski relation will  eventually (at sufficiently low temperatures) yield an overestimate of the free energy of the glass.
In a non-glassy system, all TI runs will give the same result (apart from statistical fluctuations), and the JR and TI estimates should agree.}
In section 3 of the SM, we also show that the Jarzynski methods yields  lower estimates of the free energy than the so-called basin volume method of ref.~\cite{vinutha2020numerical}.

\begin{figure}[h!]
\includegraphics[scale=0.4]{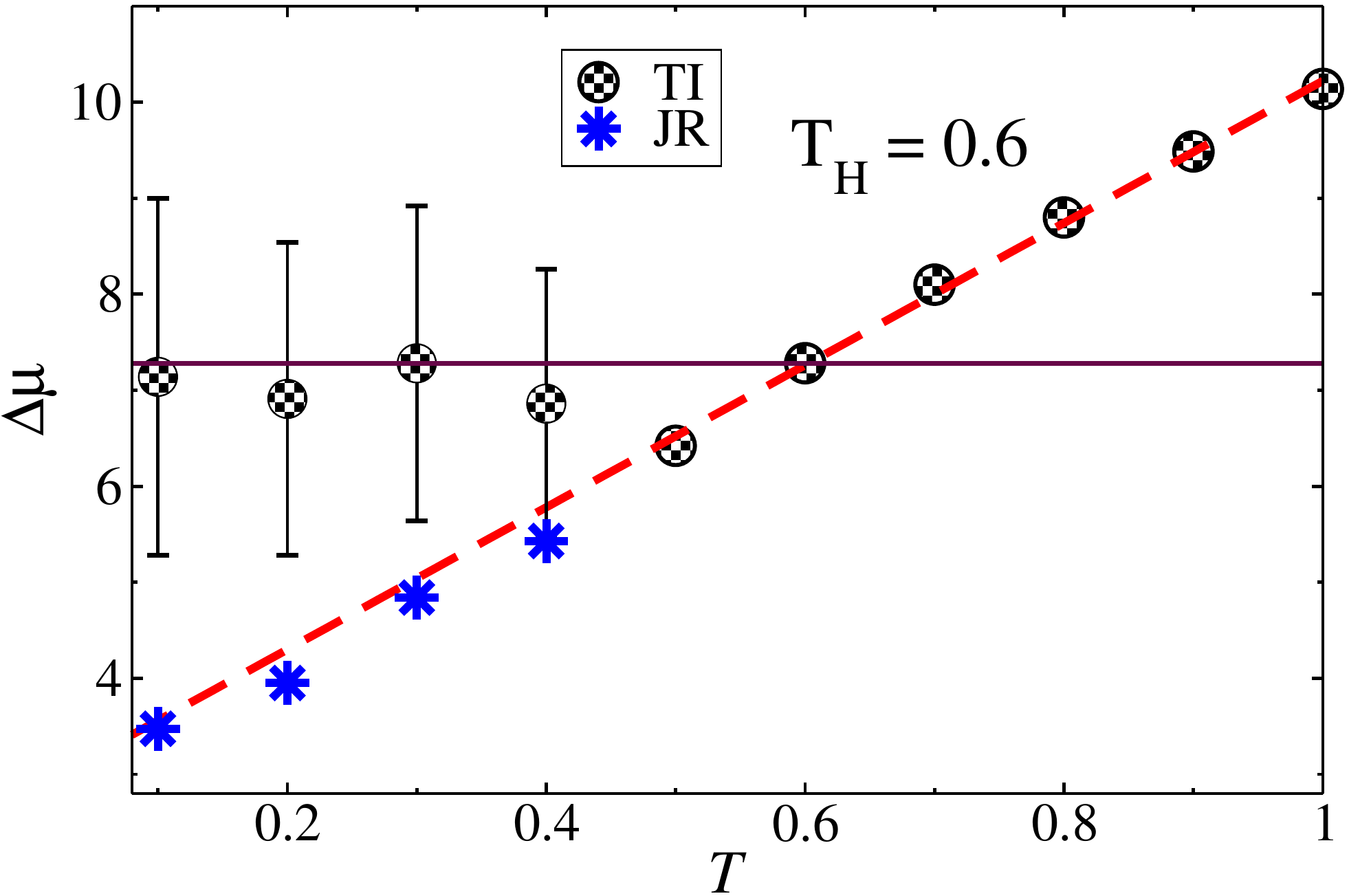}
\caption{\label{delmu} The difference in the chemical potential of the two components computed using the thermodynamic integration (TI) for the equilibrated liquid configurations at temperatures $T=1.0$ to $0.5$. 
The dashed line is a linear fit to the supercooled-liquid data. $\Delta \mu$ is computed using Jarzynski relation (stars) and TI (checkered circles) for  glassy configuration between $T = 0.1$ and $0.4$. 
The average $\Delta \mu$ computed using TI for low-$T$ glasses is close to the average $\Delta \mu$ of the liquid configurations at $T_H=0.6$ (horizontal bold line). The error bars correspond to the standard deviation of $\Delta \mu$ values.}
\end{figure}
As a separate test of the method, we also compute the chemical potential difference between the components of the KA glass, using the Jarzynski relation. 
In the case of dense KA liquids, computational techniques to probe the chemical potential, such as the Widom particle-insertion method or even the Widom particle-swap method~\cite{frenkel2001understanding}  will not work in practice.  
Rather, we used a method where we gradually transform the interaction potential of a particle from type $B$ to type $A$~\cite{mon1985chemical} and performed thermodynamic integration to estimate the chemical potential difference $\Delta \mu$ \cite{vinutha2021computation}. 
For a system in equilibrium,  this method can be used at any density and temperature.  
In our case, below $T<0.5$, the system no longer relaxes,  even during the longest simulation runs.  
In the range $T=0.1-0.4$, we use the Jarzynski relation to estimate $\Delta \mu$. 
For the low-$T$ glasses, we performed NVT MC simulations starting with the initial configurations in one of the basins which are obtained by quenching liquid configuration at $T_H = 0.6$.
We obtain the non-equilibrium work $W(t_s)$ needed to transform a B-type particle to an A-type particle from the thermodynamic integration of interaction parameters \cite{vinutha2021computation}. 
For a glassy configuration, we pick different B-type particles and transform them to an A-type particle to obtain the average amount of work done. 
We can average over different low-T glass configurations and compute $\overline{\exp(-\beta W(t_s))}$. 
Then, from Eq. \ref{eq:JR}, we obtain an estimate of $\Delta F$ for the glass configurations. 
The number of samples is ${97, 67, 66, 68}$ for temperatures ${0.1, 0.2, 0.3, 0.4}$, respectively. 
Again, Fig.~\ref{delmu} shows that the chemical potential differences estimated using the thermodynamic integration method show a wide scatter with an average value well above the extrapolated value of the supercooled liquid \cite{vinutha2021computation}. 
In contrast,  the Jarzynski method matches well with the extrapolated value of the supercooled liquid~\cite{note}. \textcolor{black}{We also computed the average $\Delta \mu$ for glasses at $T=0.1$ starting with $T_H=1.0$. Preliminary results suggested that  the average $\Delta \mu$  obtained using JR is independent of $T_H$. Additional simulations are required to systematically study the dependence of $F_{JR}$ on the cooling protocol to obtain low-T glasses.}
\\\\
We have shown that the Jarzynski method offers a a powerful tool to estimate the equilibrium properties of glasses. 
We illustrate this by using the non-equilibrium free energy expression due to Jarzynski to compute equilibrium free energies for glassy structures obtained using different cooling rates and the chemical potential difference between the components of the Kob-Andersen glass. 
The present results are of broader interest because, to our knowledge, there are no earlier examples where the Jarzynski method massively outperforms conventional free-energy calculation methods.

We expect that our work will provide a new tool to probe  the physics of amorphous solids, such as gels, glasses and jammed packings, prepared using different non-equilibrium protocols.  

\section*{Supplementary Material}
In the supplementary material (SM), we present supporting data and discussion. In SM section 1, cooling as work; section 2, work distributions; section 3, comparison between the Jarzynski and basin volume method.
\begin{acknowledgments}
We gratefully acknowledge the funding by the International
Young Scientist Fellowship of Institute of Physics (IoP), the Chinese Academy of Sciences under Grant No. 2018008. We gratefully acknowledge IoP and the University of Cambridge for computational resources and support.
\end{acknowledgments}
\section*{Data availability}
The data that support the findings of this study are
openly available in the University of Cambridge repository at
https://www.repository.cam.ac.uk/.
\\

\bibliography{JR_freeenergy_glass_v2.bib}

\begin{thebibliography}{25}%
\makeatletter
\providecommand \@ifxundefined [1]{%
 \@ifx{#1\undefined}
}%
\providecommand \@ifnum [1]{%
 \ifnum #1\expandafter \@firstoftwo
 \else \expandafter \@secondoftwo
 \fi
}%
\providecommand \@ifx [1]{%
 \ifx #1\expandafter \@firstoftwo
 \else \expandafter \@secondoftwo
 \fi
}%
\providecommand \natexlab [1]{#1}%
\providecommand \enquote  [1]{``#1''}%
\providecommand \bibnamefont  [1]{#1}%
\providecommand \bibfnamefont [1]{#1}%
\providecommand \citenamefont [1]{#1}%
\providecommand \href@noop [0]{\@secondoftwo}%
\providecommand \href [0]{\begingroup \@sanitize@url \@href}%
\providecommand \@href[1]{\@@startlink{#1}\@@href}%
\providecommand \@@href[1]{\endgroup#1\@@endlink}%
\providecommand \@sanitize@url [0]{\catcode `\\12\catcode `\$12\catcode
  `\&12\catcode `\#12\catcode `\^12\catcode `\_12\catcode `\%12\relax}%
\providecommand \@@startlink[1]{}%
\providecommand \@@endlink[0]{}%
\providecommand \url  [0]{\begingroup\@sanitize@url \@url }%
\providecommand \@url [1]{\endgroup\@href {#1}{\urlprefix }}%
\providecommand \urlprefix  [0]{URL }%
\providecommand \Eprint [0]{\href }%
\providecommand \doibase [0]{http://dx.doi.org/}%
\providecommand \selectlanguage [0]{\@gobble}%
\providecommand \bibinfo  [0]{\@secondoftwo}%
\providecommand \bibfield  [0]{\@secondoftwo}%
\providecommand \translation [1]{[#1]}%
\providecommand \BibitemOpen [0]{}%
\providecommand \bibitemStop [0]{}%
\providecommand \bibitemNoStop [0]{.\EOS\space}%
\providecommand \EOS [0]{\spacefactor3000\relax}%
\providecommand \BibitemShut  [1]{\csname bibitem#1\endcsname}%
\let\auto@bib@innerbib\@empty
\bibitem [{\citenamefont
  {Jarzynski}(1997{\natexlab{a}})}]{jarzynski1997equilibrium}%
  \BibitemOpen
  \bibfield  {author} {\bibinfo {author} {\bibfnamefont {C.}~\bibnamefont
  {Jarzynski}},\ }\bibfield  {title} {\enquote {\bibinfo {title} {Equilibrium
  free-energy differences from nonequilibrium measurements: A master-equation
  approach},}\ }\href@noop {} {\bibfield  {journal} {\bibinfo  {journal}
  {Physical Review E}\ }\textbf {\bibinfo {volume} {56}},\ \bibinfo {pages}
  {5018} (\bibinfo {year} {1997}{\natexlab{a}})}\BibitemShut {NoStop}%
\bibitem [{\citenamefont
  {Jarzynski}(1997{\natexlab{b}})}]{jarzynski1997nonequilibrium}%
  \BibitemOpen
  \bibfield  {author} {\bibinfo {author} {\bibfnamefont {C.}~\bibnamefont
  {Jarzynski}},\ }\bibfield  {title} {\enquote {\bibinfo {title}
  {Nonequilibrium equality for free energy differences},}\ }\href@noop {}
  {\bibfield  {journal} {\bibinfo  {journal} {Physical Review Letters}\
  }\textbf {\bibinfo {volume} {78}},\ \bibinfo {pages} {2690} (\bibinfo {year}
  {1997}{\natexlab{b}})}\BibitemShut {NoStop}%
\bibitem [{\citenamefont {Crooks}(1999)}]{crooks1999entropy}%
  \BibitemOpen
  \bibfield  {author} {\bibinfo {author} {\bibfnamefont {G.~E.}\ \bibnamefont
  {Crooks}},\ }\bibfield  {title} {\enquote {\bibinfo {title} {Entropy
  production fluctuation theorem and the nonequilibrium work relation for free
  energy differences},}\ }\href@noop {} {\bibfield  {journal} {\bibinfo
  {journal} {Physical Review E}\ }\textbf {\bibinfo {volume} {60}},\ \bibinfo
  {pages} {2721} (\bibinfo {year} {1999})}\BibitemShut {NoStop}%
\bibitem [{\citenamefont {Crooks}(2000)}]{crooks2000path}%
  \BibitemOpen
  \bibfield  {author} {\bibinfo {author} {\bibfnamefont {G.~E.}\ \bibnamefont
  {Crooks}},\ }\bibfield  {title} {\enquote {\bibinfo {title} {Path-ensemble
  averages in systems driven far from equilibrium},}\ }\href@noop {} {\bibfield
   {journal} {\bibinfo  {journal} {Physical review E}\ }\textbf {\bibinfo
  {volume} {61}},\ \bibinfo {pages} {2361} (\bibinfo {year}
  {2000})}\BibitemShut {NoStop}%
\bibitem [{\citenamefont {Crooks}(1998)}]{crooks1998nonequilibrium}%
  \BibitemOpen
  \bibfield  {author} {\bibinfo {author} {\bibfnamefont {G.~E.}\ \bibnamefont
  {Crooks}},\ }\bibfield  {title} {\enquote {\bibinfo {title} {Nonequilibrium
  measurements of free energy differences for microscopically reversible
  markovian systems},}\ }\href@noop {} {\bibfield  {journal} {\bibinfo
  {journal} {Journal of Statistical Physics}\ }\textbf {\bibinfo {volume}
  {90}},\ \bibinfo {pages} {1481--1487} (\bibinfo {year} {1998})}\BibitemShut
  {NoStop}%
\bibitem [{\citenamefont {Collin}\ \emph {et~al.}(2005)\citenamefont {Collin},
  \citenamefont {Ritort}, \citenamefont {Jarzynski}, \citenamefont {Smith},
  \citenamefont {Tinoco},\ and\ \citenamefont
  {Bustamante}}]{collin2005verification}%
  \BibitemOpen
  \bibfield  {author} {\bibinfo {author} {\bibfnamefont {D.}~\bibnamefont
  {Collin}}, \bibinfo {author} {\bibfnamefont {F.}~\bibnamefont {Ritort}},
  \bibinfo {author} {\bibfnamefont {C.}~\bibnamefont {Jarzynski}}, \bibinfo
  {author} {\bibfnamefont {S.~B.}\ \bibnamefont {Smith}}, \bibinfo {author}
  {\bibfnamefont {I.}~\bibnamefont {Tinoco}}, \ and\ \bibinfo {author}
  {\bibfnamefont {C.}~\bibnamefont {Bustamante}},\ }\bibfield  {title}
  {\enquote {\bibinfo {title} {Verification of the crooks fluctuation theorem
  and recovery of rna folding free energies},}\ }\href@noop {} {\bibfield
  {journal} {\bibinfo  {journal} {Nature}\ }\textbf {\bibinfo {volume} {437}},\
  \bibinfo {pages} {231--234} (\bibinfo {year} {2005})}\BibitemShut {NoStop}%
\bibitem [{\citenamefont {Douarche}\ \emph {et~al.}(2005)\citenamefont
  {Douarche}, \citenamefont {Ciliberto}, \citenamefont {Petrosyan},\ and\
  \citenamefont {Rabbiosi}}]{douarche2005experimental}%
  \BibitemOpen
  \bibfield  {author} {\bibinfo {author} {\bibfnamefont {F.}~\bibnamefont
  {Douarche}}, \bibinfo {author} {\bibfnamefont {S.}~\bibnamefont {Ciliberto}},
  \bibinfo {author} {\bibfnamefont {A.}~\bibnamefont {Petrosyan}}, \ and\
  \bibinfo {author} {\bibfnamefont {I.}~\bibnamefont {Rabbiosi}},\ }\bibfield
  {title} {\enquote {\bibinfo {title} {An experimental test of the jarzynski
  equality in a mechanical experiment},}\ }\href@noop {} {\bibfield  {journal}
  {\bibinfo  {journal} {EPL (Europhysics Letters)}\ }\textbf {\bibinfo {volume}
  {70}},\ \bibinfo {pages} {593} (\bibinfo {year} {2005})}\BibitemShut
  {NoStop}%
\bibitem [{\citenamefont {Toyabe}\ \emph {et~al.}(2010)\citenamefont {Toyabe},
  \citenamefont {Sagawa}, \citenamefont {Ueda}, \citenamefont {Muneyuki},\ and\
  \citenamefont {Sano}}]{toyabe2010experimental}%
  \BibitemOpen
  \bibfield  {author} {\bibinfo {author} {\bibfnamefont {S.}~\bibnamefont
  {Toyabe}}, \bibinfo {author} {\bibfnamefont {T.}~\bibnamefont {Sagawa}},
  \bibinfo {author} {\bibfnamefont {M.}~\bibnamefont {Ueda}}, \bibinfo {author}
  {\bibfnamefont {E.}~\bibnamefont {Muneyuki}}, \ and\ \bibinfo {author}
  {\bibfnamefont {M.}~\bibnamefont {Sano}},\ }\bibfield  {title} {\enquote
  {\bibinfo {title} {Experimental demonstration of information-to-energy
  conversion and validation of the generalized jarzynski equality},}\
  }\href@noop {} {\bibfield  {journal} {\bibinfo  {journal} {Nature physics}\
  }\textbf {\bibinfo {volume} {6}},\ \bibinfo {pages} {988--992} (\bibinfo
  {year} {2010})}\BibitemShut {NoStop}%
\bibitem [{\citenamefont {Dellago}\ and\ \citenamefont
  {Hummer}(2014)}]{dellago2014computing}%
  \BibitemOpen
  \bibfield  {author} {\bibinfo {author} {\bibfnamefont {C.}~\bibnamefont
  {Dellago}}\ and\ \bibinfo {author} {\bibfnamefont {G.}~\bibnamefont
  {Hummer}},\ }\bibfield  {title} {\enquote {\bibinfo {title} {Computing
  equilibrium free energies using non-equilibrium molecular dynamics},}\
  }\href@noop {} {\bibfield  {journal} {\bibinfo  {journal} {Entropy}\ }\textbf
  {\bibinfo {volume} {16}},\ \bibinfo {pages} {41--61} (\bibinfo {year}
  {2014})}\BibitemShut {NoStop}%
\bibitem [{\citenamefont {Lechner}\ and\ \citenamefont
  {Dellago}(2007)}]{lechner2007efficiency}%
  \BibitemOpen
  \bibfield  {author} {\bibinfo {author} {\bibfnamefont {W.}~\bibnamefont
  {Lechner}}\ and\ \bibinfo {author} {\bibfnamefont {C.}~\bibnamefont
  {Dellago}},\ }\bibfield  {title} {\enquote {\bibinfo {title} {On the
  efficiency of path sampling methods for the calculation of free energies from
  non-equilibrium simulations},}\ }\href@noop {} {\bibfield  {journal}
  {\bibinfo  {journal} {Journal of Statistical Mechanics: Theory and
  Experiment}\ }\textbf {\bibinfo {volume} {2007}},\ \bibinfo {pages} {P04001}
  (\bibinfo {year} {2007})}\BibitemShut {NoStop}%
\bibitem [{\citenamefont {Oberhofer}\ and\ \citenamefont
  {Dellago}(2009)}]{oberhofer2009efficient}%
  \BibitemOpen
  \bibfield  {author} {\bibinfo {author} {\bibfnamefont {H.}~\bibnamefont
  {Oberhofer}}\ and\ \bibinfo {author} {\bibfnamefont {C.}~\bibnamefont
  {Dellago}},\ }\bibfield  {title} {\enquote {\bibinfo {title} {Efficient
  extraction of free energy profiles from nonequilibrium experiments},}\
  }\href@noop {} {\bibfield  {journal} {\bibinfo  {journal} {Journal of
  computational chemistry}\ }\textbf {\bibinfo {volume} {30}},\ \bibinfo
  {pages} {1726--1736} (\bibinfo {year} {2009})}\BibitemShut {NoStop}%
\bibitem [{\citenamefont {Geiger}\ and\ \citenamefont
  {Dellago}(2010)}]{geiger2010optimum}%
  \BibitemOpen
  \bibfield  {author} {\bibinfo {author} {\bibfnamefont {P.}~\bibnamefont
  {Geiger}}\ and\ \bibinfo {author} {\bibfnamefont {C.}~\bibnamefont
  {Dellago}},\ }\bibfield  {title} {\enquote {\bibinfo {title} {Optimum
  protocol for fast-switching free-energy calculations},}\ }\href@noop {}
  {\bibfield  {journal} {\bibinfo  {journal} {Physical Review E}\ }\textbf
  {\bibinfo {volume} {81}},\ \bibinfo {pages} {021127} (\bibinfo {year}
  {2010})}\BibitemShut {NoStop}%
\bibitem [{\citenamefont {Angell}(1995)}]{angell1995formation}%
  \BibitemOpen
  \bibfield  {author} {\bibinfo {author} {\bibfnamefont {C.~A.}\ \bibnamefont
  {Angell}},\ }\bibfield  {title} {\enquote {\bibinfo {title} {Formation of
  glasses from liquids and biopolymers},}\ }\href@noop {} {\bibfield  {journal}
  {\bibinfo  {journal} {Science}\ }\textbf {\bibinfo {volume} {267}},\ \bibinfo
  {pages} {1924--1935} (\bibinfo {year} {1995})}\BibitemShut {NoStop}%
\bibitem [{\citenamefont {Debenedetti}\ and\ \citenamefont
  {Stillinger}(2001)}]{debenedetti2001supercooled}%
  \BibitemOpen
  \bibfield  {author} {\bibinfo {author} {\bibfnamefont {P.~G.}\ \bibnamefont
  {Debenedetti}}\ and\ \bibinfo {author} {\bibfnamefont {F.~H.}\ \bibnamefont
  {Stillinger}},\ }\bibfield  {title} {\enquote {\bibinfo {title} {Supercooled
  liquids and the glass transition},}\ }\href@noop {} {\bibfield  {journal}
  {\bibinfo  {journal} {Nature}\ }\textbf {\bibinfo {volume} {410}},\ \bibinfo
  {pages} {259--267} (\bibinfo {year} {2001})}\BibitemShut {NoStop}%
\bibitem [{\citenamefont {Berthier}\ \emph {et~al.}(2017)\citenamefont
  {Berthier}, \citenamefont {Charbonneau}, \citenamefont {Coslovich},
  \citenamefont {Ninarello}, \citenamefont {Ozawa},\ and\ \citenamefont
  {Yaida}}]{ber171}%
  \BibitemOpen
  \bibfield  {author} {\bibinfo {author} {\bibfnamefont {L.}~\bibnamefont
  {Berthier}}, \bibinfo {author} {\bibfnamefont {P.}~\bibnamefont
  {Charbonneau}}, \bibinfo {author} {\bibfnamefont {D.}~\bibnamefont
  {Coslovich}}, \bibinfo {author} {\bibfnamefont {A.}~\bibnamefont
  {Ninarello}}, \bibinfo {author} {\bibfnamefont {M.}~\bibnamefont {Ozawa}}, \
  and\ \bibinfo {author} {\bibfnamefont {S.}~\bibnamefont {Yaida}},\ }\bibfield
   {title} {\enquote {\bibinfo {title} {Configurational entropy measurements in
  extremely supercooled liquids that break the glass ceiling},}\ }\href
  {\doibase 10.1073/pnas.1706860114} {\bibfield  {journal} {\bibinfo  {journal}
  {Proceedings of the National Academy of Sciences of the United States of
  America}\ }\textbf {\bibinfo {volume} {114}},\ \bibinfo {pages}
  {11356--11361} (\bibinfo {year} {2017})}\BibitemShut {NoStop}%
\bibitem [{\citenamefont {Parmar}, \citenamefont {Ozawa},\ and\ \citenamefont
  {Berthier}(2020)}]{ADSparmar}%
  \BibitemOpen
  \bibfield  {author} {\bibinfo {author} {\bibfnamefont {A.~D.~S.}\
  \bibnamefont {Parmar}}, \bibinfo {author} {\bibfnamefont {M.}~\bibnamefont
  {Ozawa}}, \ and\ \bibinfo {author} {\bibfnamefont {L.}~\bibnamefont
  {Berthier}},\ }\bibfield  {title} {\enquote {\bibinfo {title} {Ultrastable
  metallic glasses in silico},}\ }\href {\doibase
  10.1103/PhysRevLett.125.085505} {\bibfield  {journal} {\bibinfo  {journal}
  {Phys. Rev. Lett.}\ }\textbf {\bibinfo {volume} {125}},\ \bibinfo {pages}
  {085505} (\bibinfo {year} {2020})}\BibitemShut {NoStop}%
\bibitem [{\citenamefont {Kob}\ and\ \citenamefont
  {Andersen}(1995)}]{kob1995testing}%
  \BibitemOpen
  \bibfield  {author} {\bibinfo {author} {\bibfnamefont {W.}~\bibnamefont
  {Kob}}\ and\ \bibinfo {author} {\bibfnamefont {H.~C.}\ \bibnamefont
  {Andersen}},\ }\bibfield  {title} {\enquote {\bibinfo {title} {Testing
  mode-coupling theory for a supercooled binary lennard-jones mixture i: The
  van hove correlation function},}\ }\href@noop {} {\bibfield  {journal}
  {\bibinfo  {journal} {Physical Review E}\ }\textbf {\bibinfo {volume} {51}},\
  \bibinfo {pages} {4626} (\bibinfo {year} {1995})}\BibitemShut {NoStop}%
\bibitem [{\citenamefont {Westergren}\ \emph {et~al.}(2007)\citenamefont
  {Westergren}, \citenamefont {Lindfors}, \citenamefont {H{\"o}glund},
  \citenamefont {L{\"u}der}, \citenamefont {Nordholm},\ and\ \citenamefont
  {Kjellander}}]{westergren2007silico}%
  \BibitemOpen
  \bibfield  {author} {\bibinfo {author} {\bibfnamefont {J.}~\bibnamefont
  {Westergren}}, \bibinfo {author} {\bibfnamefont {L.}~\bibnamefont
  {Lindfors}}, \bibinfo {author} {\bibfnamefont {T.}~\bibnamefont
  {H{\"o}glund}}, \bibinfo {author} {\bibfnamefont {K.}~\bibnamefont
  {L{\"u}der}}, \bibinfo {author} {\bibfnamefont {S.}~\bibnamefont {Nordholm}},
  \ and\ \bibinfo {author} {\bibfnamefont {R.}~\bibnamefont {Kjellander}},\
  }\bibfield  {title} {\enquote {\bibinfo {title} {In silico prediction of drug
  solubility: 1. free energy of hydration},}\ }\href@noop {} {\bibfield
  {journal} {\bibinfo  {journal} {The Journal of Physical Chemistry B}\
  }\textbf {\bibinfo {volume} {111}},\ \bibinfo {pages} {1872--1882} (\bibinfo
  {year} {2007})}\BibitemShut {NoStop}%
\bibitem [{\citenamefont {Sastry}(2001)}]{sastry2001relationship}%
  \BibitemOpen
  \bibfield  {author} {\bibinfo {author} {\bibfnamefont {S.}~\bibnamefont
  {Sastry}},\ }\bibfield  {title} {\enquote {\bibinfo {title} {The relationship
  between fragility, configurational entropy and the potential energy landscape
  of glass-forming liquids},}\ }\href@noop {} {\bibfield  {journal} {\bibinfo
  {journal} {Nature}\ }\textbf {\bibinfo {volume} {409}},\ \bibinfo {pages}
  {164--167} (\bibinfo {year} {2001})}\BibitemShut {NoStop}%
\bibitem [{\citenamefont {Sengupta}\ \emph {et~al.}(2011)\citenamefont
  {Sengupta}, \citenamefont {Vasconcelos}, \citenamefont {Affouard},\ and\
  \citenamefont {Sastry}}]{sengupta2011dependence}%
  \BibitemOpen
  \bibfield  {author} {\bibinfo {author} {\bibfnamefont {S.}~\bibnamefont
  {Sengupta}}, \bibinfo {author} {\bibfnamefont {F.}~\bibnamefont
  {Vasconcelos}}, \bibinfo {author} {\bibfnamefont {F.}~\bibnamefont
  {Affouard}}, \ and\ \bibinfo {author} {\bibfnamefont {S.}~\bibnamefont
  {Sastry}},\ }\bibfield  {title} {\enquote {\bibinfo {title} {Dependence of
  the fragility of a glass former on the softness of interparticle
  interactions},}\ }\href@noop {} {\bibfield  {journal} {\bibinfo  {journal}
  {The Journal of chemical physics}\ }\textbf {\bibinfo {volume} {135}},\
  \bibinfo {pages} {194503} (\bibinfo {year} {2011})}\BibitemShut {NoStop}%
\bibitem [{\citenamefont {Vinutha}\ and\ \citenamefont
  {Frenkel}(2020)}]{vinutha2020numerical}%
  \BibitemOpen
  \bibfield  {author} {\bibinfo {author} {\bibfnamefont {H.~A.}\ \bibnamefont
  {Vinutha}}\ and\ \bibinfo {author} {\bibfnamefont {D.}~\bibnamefont
  {Frenkel}},\ }\bibfield  {title} {\enquote {\bibinfo {title} {Numerical
  method for computing the free energy of glasses},}\ }\href@noop {} {\bibfield
   {journal} {\bibinfo  {journal} {Physical Review E}\ }\textbf {\bibinfo
  {volume} {102}},\ \bibinfo {pages} {063303} (\bibinfo {year}
  {2020})}\BibitemShut {NoStop}%
\bibitem [{\citenamefont {Frenkel}\ and\ \citenamefont
  {Smit}(2001)}]{frenkel2001understanding}%
  \BibitemOpen
  \bibfield  {author} {\bibinfo {author} {\bibfnamefont {D.}~\bibnamefont
  {Frenkel}}\ and\ \bibinfo {author} {\bibfnamefont {B.}~\bibnamefont {Smit}},\
  }\href@noop {} {\emph {\bibinfo {title} {Understanding molecular simulation:
  from algorithms to applications}}},\ Vol.~\bibinfo {volume} {1}\ (\bibinfo
  {publisher} {Elsevier},\ \bibinfo {year} {2001})\BibitemShut {NoStop}%
\bibitem [{\citenamefont {Mon}\ and\ \citenamefont
  {Griffiths}(1985)}]{mon1985chemical}%
  \BibitemOpen
  \bibfield  {author} {\bibinfo {author} {\bibfnamefont {K.}~\bibnamefont
  {Mon}}\ and\ \bibinfo {author} {\bibfnamefont {R.~B.}\ \bibnamefont
  {Griffiths}},\ }\bibfield  {title} {\enquote {\bibinfo {title} {Chemical
  potential by gradual insertion of a particle in monte carlo simulation},}\
  }\href@noop {} {\bibfield  {journal} {\bibinfo  {journal} {Physical Review
  A}\ }\textbf {\bibinfo {volume} {31}},\ \bibinfo {pages} {956} (\bibinfo
  {year} {1985})}\BibitemShut {NoStop}%
\bibitem [{\citenamefont {Vinutha}\ and\ \citenamefont
  {Frenkel}(2021)}]{vinutha2021computation}%
  \BibitemOpen
  \bibfield  {author} {\bibinfo {author} {\bibfnamefont {H.~A.}\ \bibnamefont
  {Vinutha}}\ and\ \bibinfo {author} {\bibfnamefont {D.}~\bibnamefont
  {Frenkel}},\ }\bibfield  {title} {\enquote {\bibinfo {title} {Computation of
  the chemical potential and solubility of amorphous solids},}\ }\href@noop {}
  {\bibfield  {journal} {\bibinfo  {journal} {J Chem Phys}\ }\textbf {\bibinfo
  {volume} {154}},\ \bibinfo {pages} {124502} (\bibinfo {year}
  {2021})}\BibitemShut {NoStop}%
\bibitem [{not()}]{note}%
  \BibitemOpen
  \href@noop {} {}\bibinfo {note} {A preliminary version of Fig. \ref{delmu},
  with poorer statistics, was reported in
  ref.~\onlinecite{vinutha2021computation}.}\BibitemShut {Stop}%
\end{thebibliography}%

\end{document}


\title{Supplementary material: Estimation of the equilibrium free energy for glasses using the Jarzynski equality}
\author{H A Vinutha and Daan Frenkel}
\date{}
\maketitle
\section{Cooling as work}
In the main text, we apply the Jarzynski relation to estimate the free energy change upon cooling.
This may seem strange, because changing the temperature is achieved by heat transfer, rather than by work.
Here we show that the free energy change upon cooling may be interpreted in terms of work.

Consider a system with a potential energy function $U(x)$, where $x$ denotes the set of $dN$ coordinates.
Then the configurational part of the free energy is:
\begin{equation}
F(T)=-k_BT\ln \int d x\; e^{-\beta U(x)}
\end{equation}
We wish to estimate $F(T^\prime)$ with $T^\prime=T/\lambda$.
Clearly,
\begin{equation}\label{eq:JR}
F(T^\prime)=-k_BT^\prime\ln \int d x\; e^{-(\beta \lambda) U(x)}
\end{equation}
We write this as
\begin{equation}
\beta^\prime F(T^\prime;U)= - \ln \int d x\; e^{-(\beta \lambda) U(x)}=
- \ln \int d x\; e^{-\beta (\lambda U(x))} = \beta F(T;\lambda U)
\end{equation}
In other words,  the scaled free energy of a system at the original temperature but with potential energy function $\lambda U(x)$ has the same value as the scaled free energy
$\beta^\prime F(T^\prime)$ for the original potential energy function, but at 
temperature $T^\prime=T/\lambda$.
For the calculation, we compute the variation of $\beta F$, as we change the potential energy at constant $\beta$ from $U(x)$ to $\lambda U(x)$.
This transformation can be viewed as mechanical work. 

In practice, we compute
\begin{equation}
\beta [F(T;\lambda U) - F(T;U)]
=
-\ln\langle e^{-\beta\int_1^\lambda d\lambda^\prime \; \overline{U}_{\lambda^\prime} }\rangle,
\end{equation}
\textcolor{black}{where $\langle...\rangle$ denotes averaging over all independent cooling runs, and $\overline{U}_{\lambda^\prime}$ denotes the average of $U_{\lambda^\prime}$  during a single cooling run.}

\section{\textcolor{black}{Work distributions}}
\begin{figure}[h!]
\includegraphics[scale=0.4,angle=0]{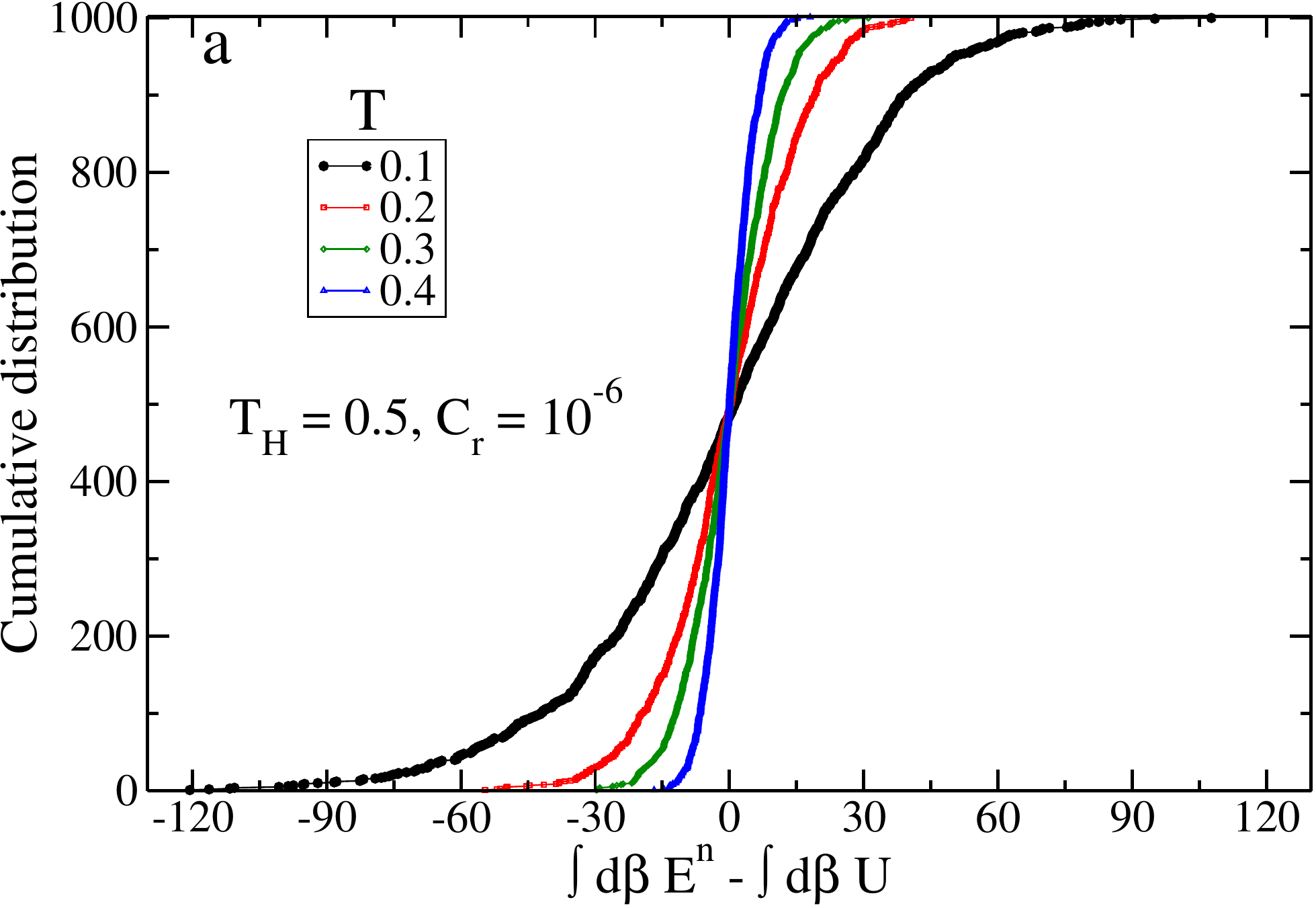}
\includegraphics[scale=0.4,angle=0]{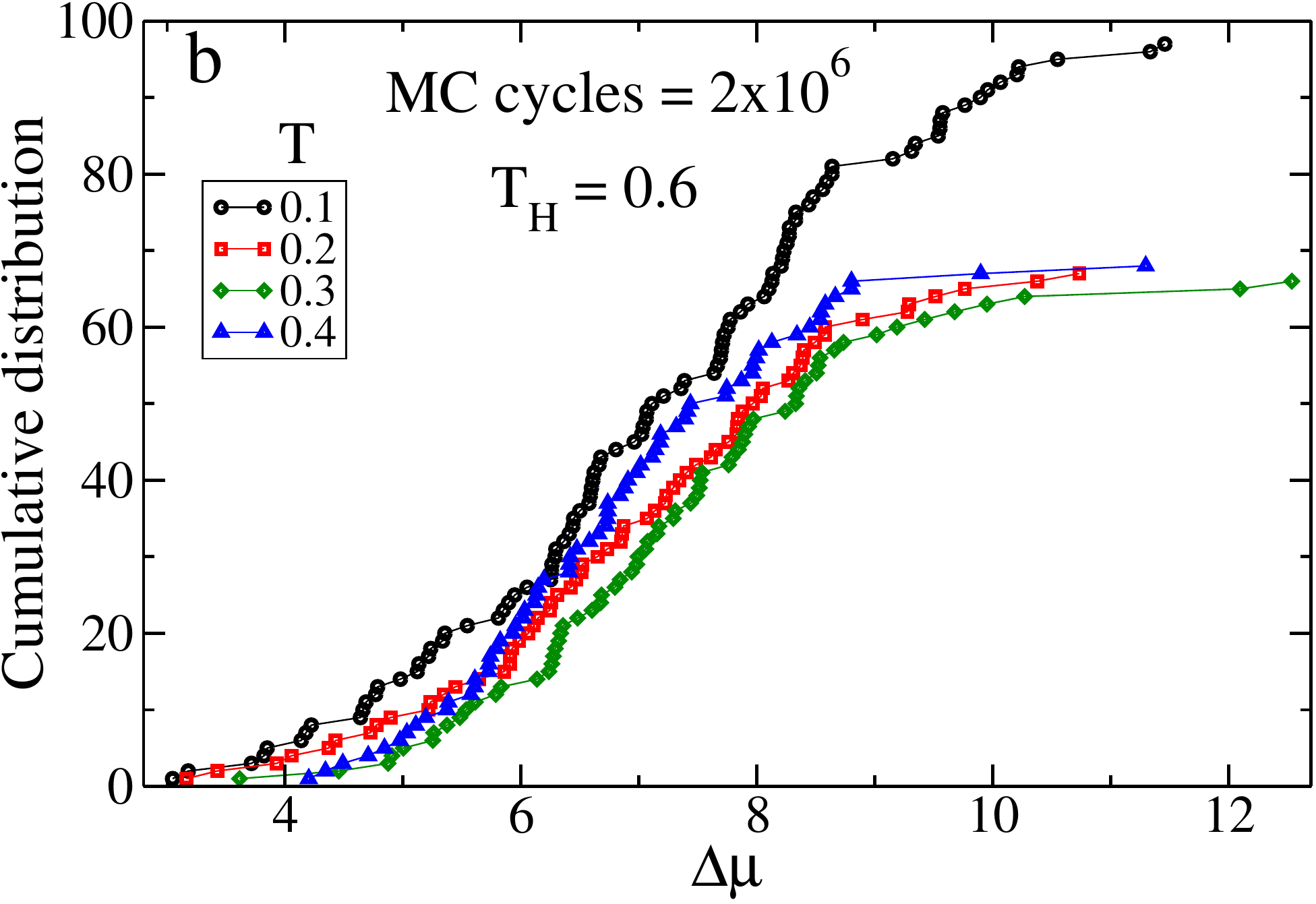}
\caption{\label{work} \textcolor{black}{ {\bf (a)} Cumulative work distributions shown for different temperatures, for the cooling runs data. {\bf(b)} Cumulative distribution of $\Delta \mu$ as a function of $T$. Each data point corresponds to the work done to transform a B-type particle to an A-type particle at that temperature.}}
\end{figure}
\textcolor{black}{In Fig. \ref{work}, we show the cumulative work distributions for the cooling runs at different temperatures and the $\Delta \mu$ calculations. 
In Fig. \ref{work}(a), we show the difference between the work done for different cooling runs and the average work done value $\int d\beta U$. 
In Fig. \ref{work}(b), we show the work needed to transform different B-type particles to A-type particles for different low-T glass configurations. 
We observe that the distribution becomes broader as the temperature decreases. 
We know that the low-energy inherent structures are dominant at low temperatures. 
The exponential weighting in Eq.~\ref{eq:JR} biases the average towards the lower energy states. 
Therefore, we obtain the lower free energy estimates from JR compared to TI. 
However, it is surprising that even at $T=0.1$, which is well below the glass transition temperature $T_g \approx 0.34$ [16], with the limited sampling we can obtain reasonable estimates of the equilibrium free energies for glasses and the average $\Delta \mu$ that matches the extrapolated supercooled value.}

\section{Comparison between the Jarzynski and basin volume method}
\begin{figure*}[h!]
\centering
\includegraphics[scale=0.35]{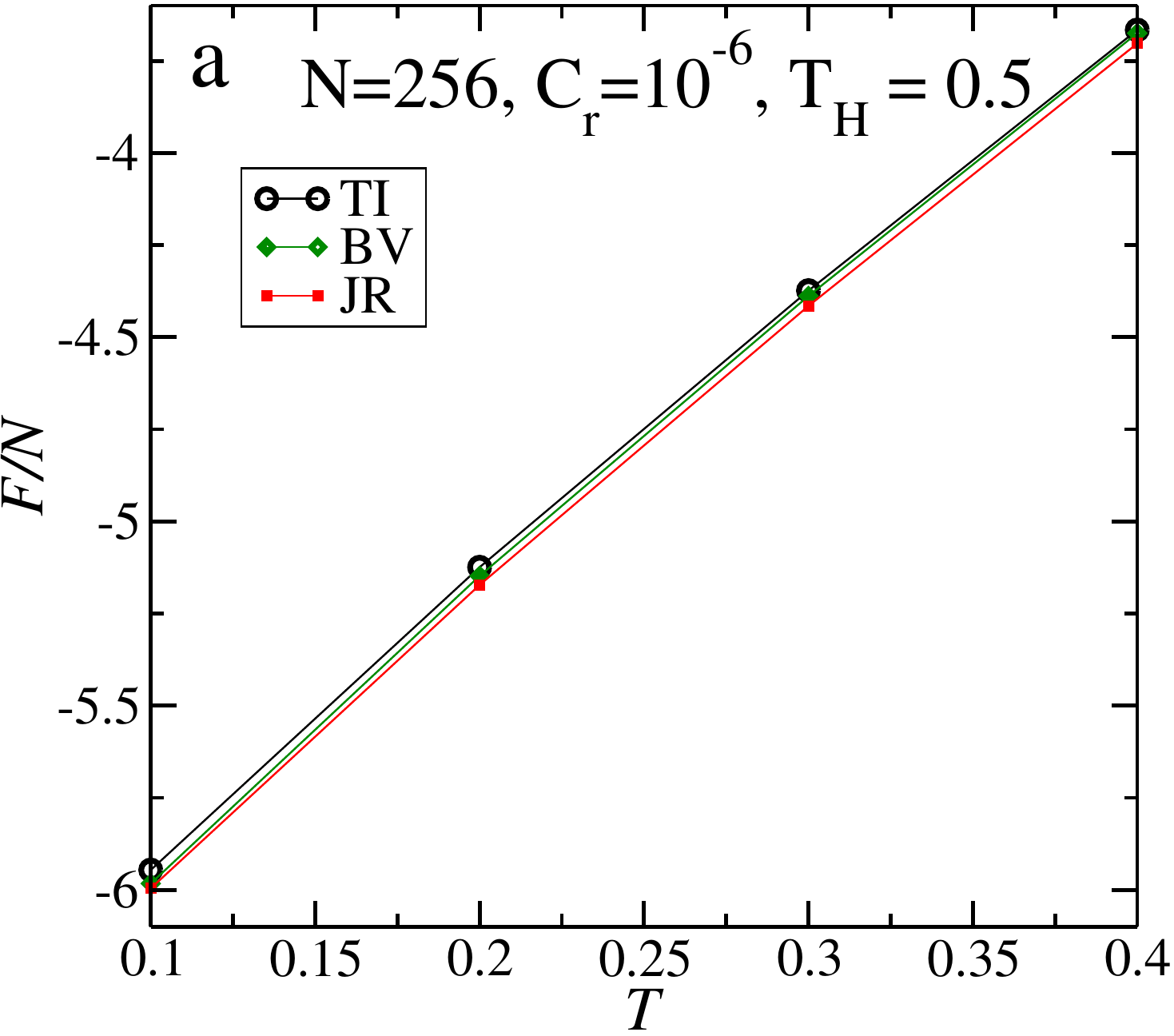}
\includegraphics[scale=0.35]{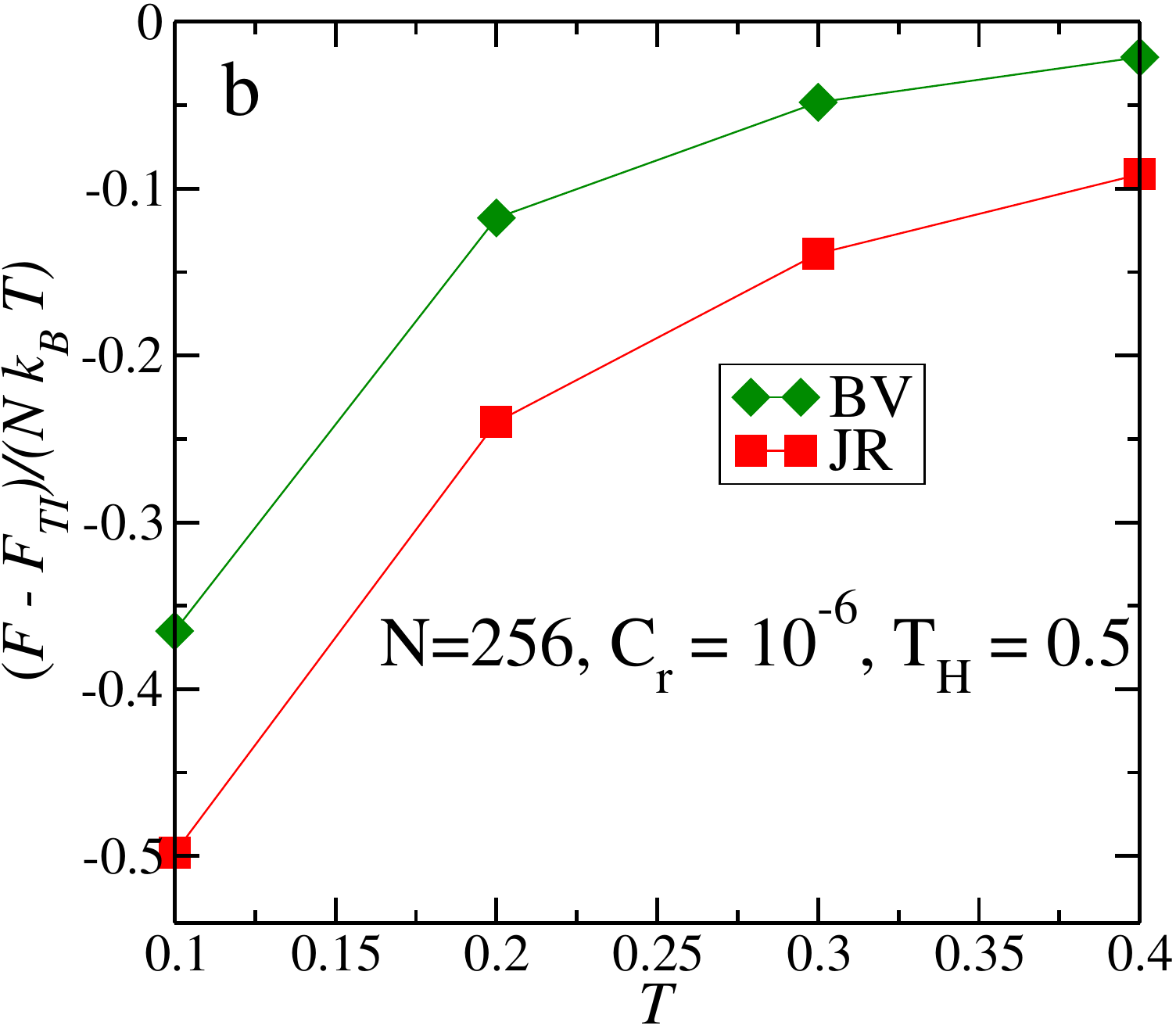}
\includegraphics[scale=0.35]{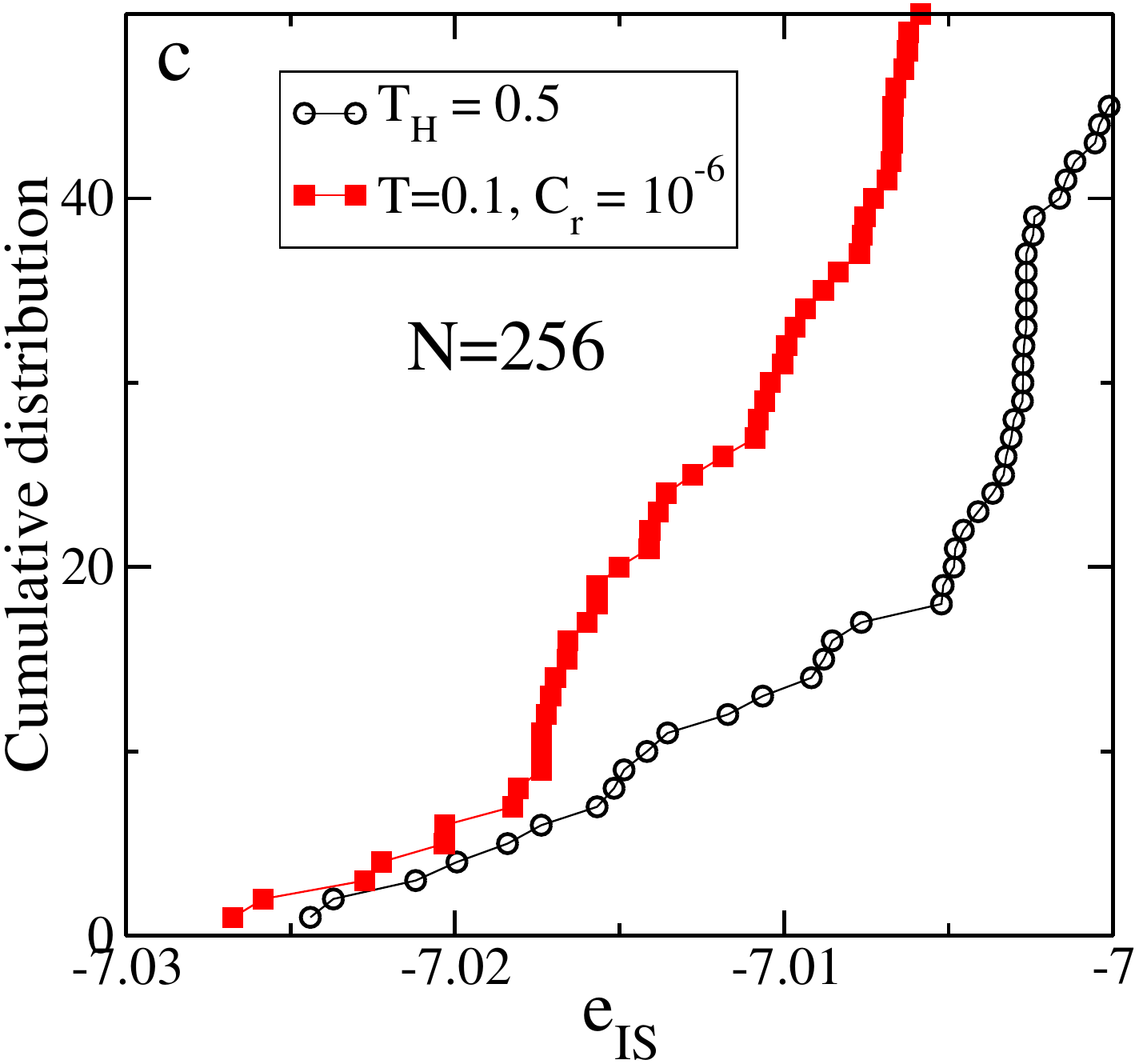}
\caption{\label{JRBV} {\bf a} Comparison between the low-T free energy obtained using TI, the basin volume (BV) method and JR. We can see the difference clearly in {\bf b}. We observe that the estimates from JR are accurate than the BV method. {\bf c} Cumulative distribution of the inherent structure energies ($e_{IS}$), shown for $T_H = 0.5$ and from the cooling runs at $T=0.1$.}
\end{figure*}
Here, we discuss the comparison between the Jarzynski method (JR) and the recently developed basin volume (BV) technique to estimate the equilibrium free energy of glassy materials [22]. In Fig. \ref{JRBV} (a), we compare the free energies computed using different methods for low-T glasses. We see in Fig. \ref{JRBV}(b) that the basin volume method performs better than TI but the Jarzynski method beats the BV and TI methods. In the BV method, we perform more than $900$ instantaneous quenches from the liquid configurations at $T_{\text{H}}=0.5$ using the conjugate gradient minimization method. The initial liquid configurations of the quench are obtained by Monte Carlo sampling and are therefore Boltzmann weighted. Now, we know the free energy at high temperature $F(T_H)$, we can obtain configurational free energy for low-temperature ($T_L$) glasses, by performing a large number of thermodynamic integrations where we cool the system confined to a given basin from $T_H$ to $T_L$, see Ref. 22 for more details. For the cooling runs, we use the same set of initial configuration $T_H$. To understand the difference in the free energy estimates from the two methods we looked at the cumulative distribution of the inherent structure energies. In Fig. \ref{JRBV}(c),for $T=0.1,C_r = 10^{-6}$, we see that the system samples lower energy inherent structures compared to the distribution at $T_H = 0.5$. The reason is that, whereas in the basin volume method, the system is constrained to remain in a single basin during cooling, transitions to lower energy basins are possible during the Jarzynski cooling protocol. 
As a consequence, the Jarzynski method achieves better sampling of the low-energy inherent structures of the system. But, of course, the Jarzynski method is also approximate.
Hence, we should expect the true equilibrium free energy of the glassy system to be even lower than the Jarzynski estimate. 

This observation also suggests that with the Jarzynski approach, it is better to carry out many simulations on a small system, rather than fewer on a large system: for TI we would not see much difference in accuracy. However, the smaller the number of runs, the smaller the chances of sampling the relevant low-energy structures.
Indeed, we found that, for the same computational effort,  the Jarzynski method yielded significantly higher free energies estimates for a system of N=1000 particles than for N=256. \textcolor{black}{Presumably, with sufficient sampling, the free energy estimates for larger systems would match the $N=256$ results. In view of the high computational costs, we did not attempt long simulations of the $N$=1000-system.}